\documentclass[11pt,twoside]{article}
\usepackage{asp2010}
\resetcounters

\begin{document}

%\warningoverprint{Draft 2010/11/30}

%\title{Solar energetic events and the solar-stellar connection}
\title{Solar energetic events, the solar-stellar connection, and
  statistics of extreme space weather}
\author{C.J. Schrijver
\affil{Lockheed Martin Advanced Technology Center, 3251 Hanover
  Street, Palo Alto, CA 94304, U.S.A.}}

\begin{abstract}
  Observations of the Sun and of Sun-like stars provide access to
  different aspects of stellar magnetic activity that, when combined,
  help us piece together a more comprehensive picture than can be
  achieved from only the solar or the stellar perspective. Where the
  Sun provides us with decent spatial resolution of, e.g., magnetic
  bipoles and the overlying dynamic, hot atmosphere, the ensemble of
  stars enables us to see rare events on at least some
  occasions. Where the Sun shows us how flux emergence, dispersal, and
  disappearance occur in the complex mix of polarities on the surface,
  only stellar observations can show us the activity of the ancient or
  future Sun. In this review, I focus on a comparison of statistical
  properties, from bipolar-region emergence to flare energies, and
  from heliospheric events to solar energetic particle
  impacts on Earth. In doing so, I point out some intriguing
  correspondences as well as areas where our knowledge falls
  short of reaching unambiguous conclusions on, for example, the most extreme
  space-weather events that we can expect from the present-day
  Sun. The difficulties of interpreting stellar coronal light curves in terms 
  of energetic events are illustrated with some examples provided by
  the SDO, STEREO, and GOES spacecraft.
\end{abstract}

\section{Introduction}
Magnetic activity of Sun and Sun-like or ``cool'' stars results in a
rich variety of observable phenomena that range from 
the asterospheres that surround these stars down to the stellar
surfaces and --~with rapid advances expected in asteroseismology~-- below. Many of
these phenomena are directly observable on Sun and stars alike,
including the outer-atmospheric phenomena of persistent chromospheres
and coronae, as well as their perturbations in the form of short-lived
light-curve perturbations that are the signature of energetic
events. It is on the latter that I focus here.

The proximity of the Sun enables us to see details in the evolving
magnetic field and associated atmospheric phenomena that are simply
impossible to infer from stellar observations. Small wonder, then,
that many of the names for stellar phenomena are taken from the solar
dictionary: active regions, spots, flares, eruptions, and even the
processes such as differential rotation and meridional advection that
are part of the equivalent of the magnetic activity cycle, all the way
to the loss of angular momentum associated with the gusty
outflow of hot, magnetized plasma. Ensembles of observations taken
over periods of years to centuries are revealing the statistical
properties of some of these phenomena, even as state-of-the art
observatories in space and on the ground are revealing physical
processes and the interconnectedness of the global outer atmosphere.

But recent observations of the Sun only provide a very limited view of what
its magnetic activity has on offer, for at least two
reasons. First, our Sun has a magnetic activity cycle that is
relatively long compared to researchers' careers as well as to the era
of advanced technology that has aided us in our
observations. Consequently, we can expect even the 'present-day Sun'
to have surprises in store for us that we have not yet observed simply
because we have not been looking long enough.  Some of these surprises
may lie hidden in records such as polar ice sheets, while others lie
embedded in rocks from outer space. But the lessons that can be
learned from these invaluable and, as yet, under-explored archives are
limited by access to these resources, by the limited
temporal resolution of such records, and by the long chains of
processes that sit between a solar phenomenon like the sunspot cycle
or its largest flares and the 'recording physics' for the 
archive from which we are attempting to learn
about them. In addition to learning about the Sun from such
'geological' records of its activity, one can also perform an ensemble
study of states of infrequent extreme 
solar activity by looking at a sample of stars like it.
This can provide us with a large
enough sample of Sun-like stars that we can begin to assess how 
frequently the Sun may subject us to rare but high-impact
events such as dangerous superflares and disruptive geomagnetic storms:
although rare, the damage that may be inflicted to our global society 
and its safety and economy by extreme events is of such a magnitude that
in-depth study of their properties and likelihood is prudent.\footnote{See the NRC report on
  ``Severe Space Weather Events -- Understanding Societal and Economic
  Impacts" at 
\hbox{http://www.nap.edu/catalog.php?record$\_$id=12507}.}

A second reason why stellar studies are crucial to understanding
of solar activity is that only stellar observations allow us to
explore what the Sun's activity has been in the very distant past or
what it will be in the very distant future (measured on time scales up to
billions of years) by selecting stars of a wide range of ages.

In this review, I discuss a sampling of the results coming out of the
study of what has been termed the solar-stellar connection. I focus,
in particular, on lessons that we are learning about what could be
called 'space climate', i.e., the characteristic state of activity of
a star like the Sun including the fluctuations about the mean in the
form of energetic events like flares and coronal mass ejections
(CMEs).  Consequently, one of the topics selected for this review is a
comparison of frequency distributions of bipolar regions, flares,
CMEs, and solar energetic particle (SEP) events, the possible relationships between them, and the lessons learned
by combining solar and stellar observations in our quest to 
establish the 'laws' of astro-magnetohydrodynamics.  Another topic is that of light curves, which touches on the need to have
pan-chromatic knowledge to guide our interpretation of stellar
observations as well as on the long-standing concept of 'sympathetic
events' in stellar magnetic activity.

\section{The flux spectrum of bipolar regions}\label{sec:bipoles}
Magnetic flux emerges from the solar interior as flux bundles that
shape themselves into bipolar regions. Those regions large enough to
contain spots during at least some of their mature, coherent phase,
are called active regions. Their emergence frequency for sunspot cycle
22 was found to be characterized by a power-law distribution with an
index $-\alpha_\Phi = 2.0\pm 0.1$ \citep{harvey+zwaan93,ref254}.

The flux bundles that are too small to contain spots, typically with
fluxes below about $10^{20}$\,Mx, but larger than about $10^{18}$\,Mx,
are called ephemeral regions. The flux spectrum of the
ephemeral regions appears to smooth transition into that of the
active region spectrum.
When the frequency distributions for active and ephemeral regions are combined
into an approximate power-law distribution, one finds
\begin{equation}\label{eq:fluxspectrum}
N(\Phi){\rm d}\Phi \propto \Phi^{-\alpha_\Phi}{\rm d}\Phi
\end{equation}
for $\alpha_\Phi=2.69$ \citep{thornton+parnell2010}. The
difference between the above values of $\alpha_\Phi$ indicates that a
single power-law is too simple as an approximation, with the flux
spectrum somewhat less steep with increasing size. For the purpose of
the discussion below, I will take the range as a measure of
uncertainty, i.e. $\alpha_\Phi = 2.3 \pm 0.3$.

The active-region spectrum for cycle 22 appeared to be roughly fixed
in shape, going up and down by a multiplicative factor over the cycle
with a power-law index changing by at most a few tenths of its value
\citep{harvey+zwaan93,ref254}. During the recent cycle minimum (late
2008 and early 2009), active regions were essentially absent for a
long time, with the 3-month average sunspot number hovering around an
unusually low value of 1.5, down from around 150 during characteristic
sunspot maxima. The ephemeral-region population, in contrast, remained
essentially unchanged \citep{schrijver+etal2011}. Hence, the
bipolar-region spectrum is likely not a single, time-independent power
law. Yet, at least during relatively active phases, on which I shall
focus below, this simple approximation appears warranted.

\begin{figure}[!ht]
%\plotone{lightcurvesgoes.ps}
%\plotone{lightcurves.ps}
\plotone{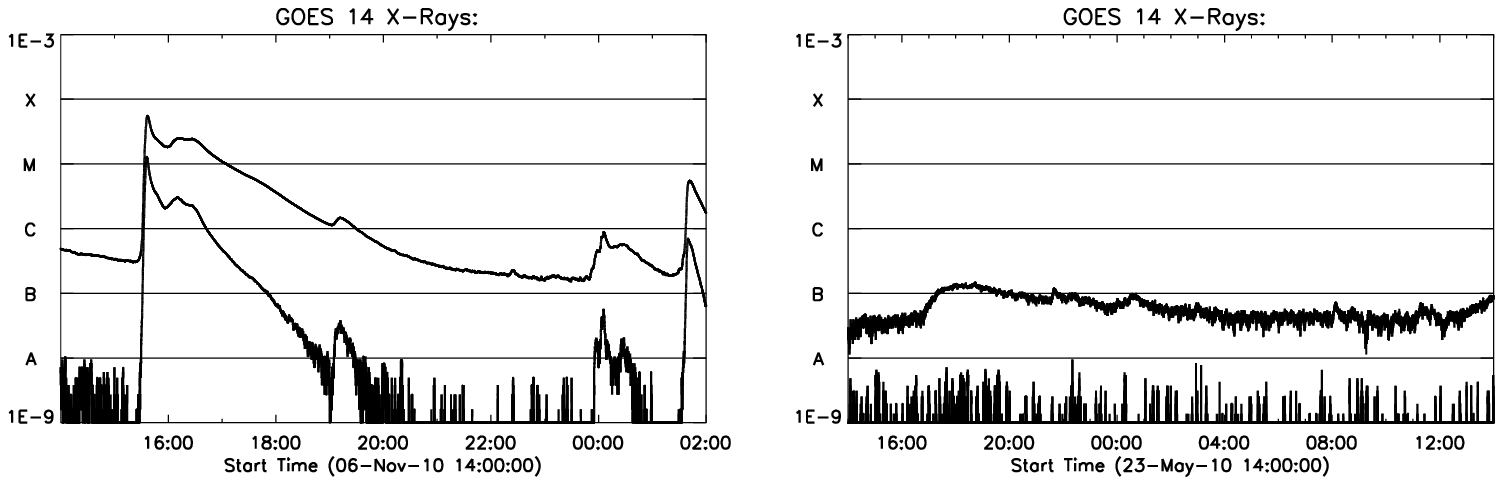}
\plotone{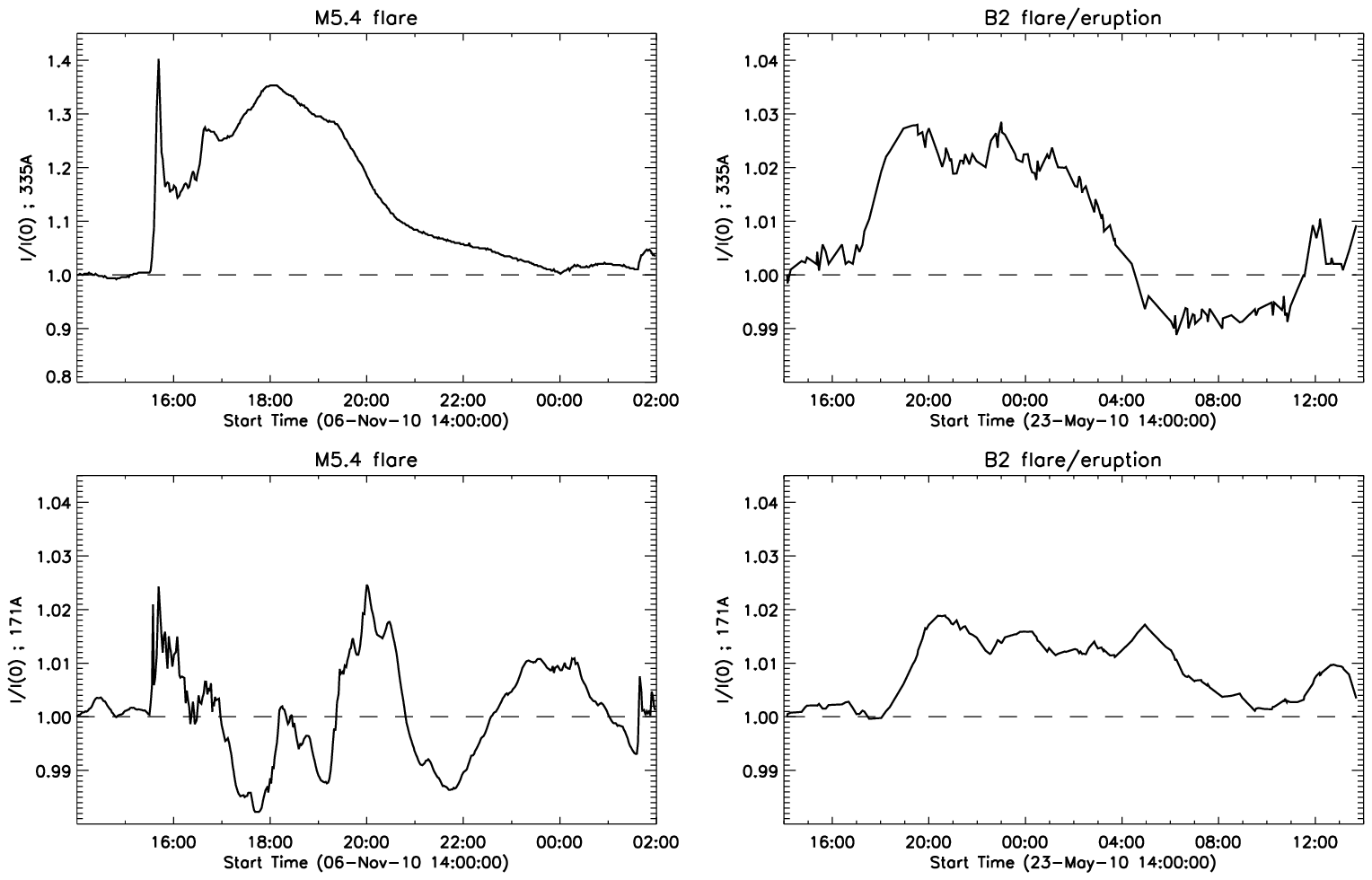}
\caption{Sample X-ray lightcurves observed with GOES (top) and EUV
  signals observed with SDO/AIA in the Fe XVI 335\,\AA\ channel
  (center) and the Fe IX/X 171\,\AA\ channel (bottom), for an M5.4
  active-region flare (left) and for the eruption of a large filament
  from a quiet-Sun region (right). The intensities were scaled to a
  pre-event reference value (at the dashed line). Note that the
  vertical scale of the center-left panel is 10$\times$ that of the
  other AIA panels. \label{fig:lightcurves}}
\end{figure}
Eq.~(\ref{eq:fluxspectrum}) characterizes the frequency of emerging
bipoles.  In a study of bipolar regions existing on the solar surface
during 1996 to 2008, \citet{zhang+etal2010} find a power law with an
index of $\alpha_{\rm exist} \sim 1.63$; its lower (absolute) value
than $\alpha_\Phi$ in Eq.~(\ref{eq:fluxspectrum}) is caused by the
longer life time of larger active regions.

The flux distribution reported on by \citet{zhang+etal2010} exhibits a
marked drop below the power law for fluxes exceeding $\Phi \sim 6 \times
10^{23}$\,Mx, and they find no regions above $\Phi_{\rm max} \sim 2
\times 10^{24}$\,Mx. Historically, the largest sunspot group recorded
occurred in April of 1946, with a value of 6\,milliHemispheres
\citep{taylor1989}; for an estimated field strength of 3\,kG, that
amounts to a flux in the spot group alone of $\Phi_{\rm spots} \sim 6
\times 10^{23}$\,Mx. The total flux in this spot group was likely
larger, but perhaps within a factor of $2-3$ of that in the spots, and
thus of the same order of magnitude as the upper limit to the
distribution found by \citet{zhang+etal2010}.

During their mature, coherent phase solar active regions are characterized
by a remarkably similar flux density, $\langle B \rangle$
of about $100$\,Mx/cm$^2$ to  $150$\,Mx/cm$^2$ \citep{ref254} regardless of region size. This
allows us to perform a transformation of the frequency distribution of
fluxes to one of total energies in the atmospheric field (to be
discussed below), using a
simple
approximation that the energy $E_B$ contained in the bipolar-region
field for an active region with size $\ell$
scales as
\begin{equation}
E_B \propto \langle B \rangle^2 \ell^3 \propto (\langle B\rangle \ell^2)^{3/2} = \Phi^{3/2}.
\end{equation}
Thus, when rewriting 
Eq.~(\ref{eq:fluxspectrum}) in terms
of energies one finds:
\begin{equation}\label{eq:freqdistflux}
N(\Phi) {\rm d}\Phi = n(E_B) {{\rm d}\Phi \over {\rm d} E_B} {\rm d} E_B \propto E_B^{-{2\over
    3}\alpha_\Phi - {1\over 3}} {\rm d}E_B\equiv
    E_B^{-\mu} {\rm d}E_B= E_B^{-1.9\pm 0.2}{\rm d}E_B.
\end{equation}
For the above value of $\Phi_{\rm max}$, the corresponding energy
is $E_{B,{\rm max}} \sim 10^{37}$\,ergs.

\section{The spectral appearance of flares and eruptions}\label{sec:spectral}
Solar flares span an astonishing range of energies: from the largest
flares, emitting brightly in hard X-rays and even $\gamma$ rays, down
to the weakest EUV flares barely detectable against the persistent background
glow of the quiet-Sun corona. The fact that flares shift in color from
EUV for $\sim 10^{24}$-erg nanoflares to hard X-rays for $\sim
10^{32}$-erg X-class events makes a statistical comparison difficult:
small and large flares are generally observed with different
instruments, exacerbating the difficulties in estimating
total energies in the absence of bolometric observables (solar flares
are too faint to be picked up by total irradiance monitors, with very
few exceptions). This is compounded
by problems in estimating
the strong background in the case of the weakest events. For the
largest flares, there are
statistical limitations owing to their low frequency.
 
The launch of the Solar Dynamics Observatory (SDO) in the spring of
2010 has opened a window onto the global Sun that enables a direct
comparison of solar and stellar observations in terms of light curves
and their interpretation.  The Atmospheric Imaging Assembly
\citep[AIA;][]{aiainstrument} on SDO is a full-disk EUV imaging
telescope array that provides high-resolution ($\sim 1.4$\,arcsec),
high-cadence ($\sim 12$\,s) coronal images in 7 narrow-band spectral
windows (in addition to a few UV channels). SDO's Extreme-ultraviolet Variability Experiment (EVE)
measures the X-ray and EUV spectral irradiance
\citep[e.g.,][]{woods+etal2010}. The comparison of the data from these
two instruments is just starting, but already some unanticipated
findings are emerging.

One of these surprises has to do with the relationship between flares
and CMEs. With the Sun in a moderately active state through much of 2010, SDO
has thus far observed primarily C-class flares and only a few low
M-class flares. \citet{woods+etal2010} note that the AIA observations
show that about 80\%\ of the C-class flares are associated with
CMEs. Earlier work had suggested numbers as low as $\sim 20$\%\ for
C-class flares, increasing to $\sim 40$\%\ and $\sim 100$\%\ for M-
and X-class flares, respectively \citep[see summary and references
in][]{schrijver2009b}. The reason for the relatively large fraction of
eruptive events remains subject to investigation; it may reflect a
selection bias in earlier studies, or may suggest that flaring during
low-activity states more readily breaks through the relatively weak
overlying field than during higher activity states.

On another front, even the Sun still has surprises in store as to how
flares show themselves in different pass bands.  Associated with the
large fraction of eruptive C-class flares is a characteristic
signature seen in the EVE spectral irradiance measurements:
\citet{woods+etal2010} point out that these eruptive flares have
late-phase emissions in, e.g., Fe~XV and Fe~XVI that AIA data show to
be associated with relatively high post-eruption loop systems,
apparently reconnecting and cooling after the eruptive flare. That
late-phase emission was not observed for C-class solar flares until
now.

Another example of what we are learning about solar energetic events
is shown in Fig.~\ref{fig:lightcurves}. The top-left panel shows the
GOES $0.5-4.0$\,\AA\ (lower curve) and $1.0-8.0$\,\AA\ (upper curve)
signals for the 2010/11/05 M5.4 flare. The AIA lightcurves below it
are for the Fe~XVI 335\,\AA\ and Fe~IX/X 171\,\AA\ channels. This
eruptive flare shows a spike in the Fe~XVI 335\,\AA\ channel followed
by a long-duration signal indicative of the cooling of post-eruption
loops (as discussed above for the less energetic C-class flares). The
much cooler Fe~IX/X 171\,\AA\ signal shows a mixture of
brightenings and darkenings that reflect heating, cooling, and
expansion-related coronal dimmings. For an analogous
discussion on stellar flaring, see, e.g., \citet{osten+brown1999}.

\begin{figure}[!ht]
%\plotone{lightcurvesgoes_20100801.ps}
%\plotone{lightcurves_20100801.ps}
\plotone{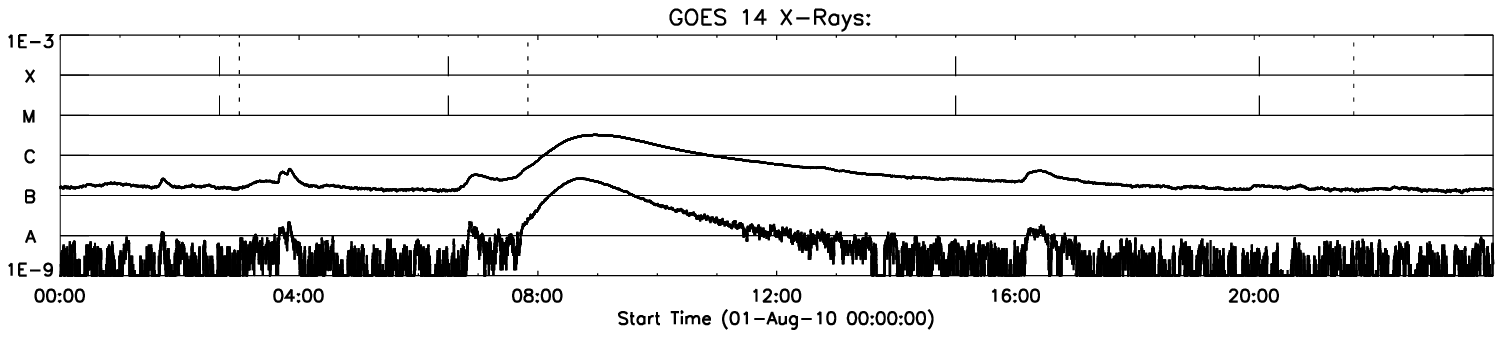}
\plotone{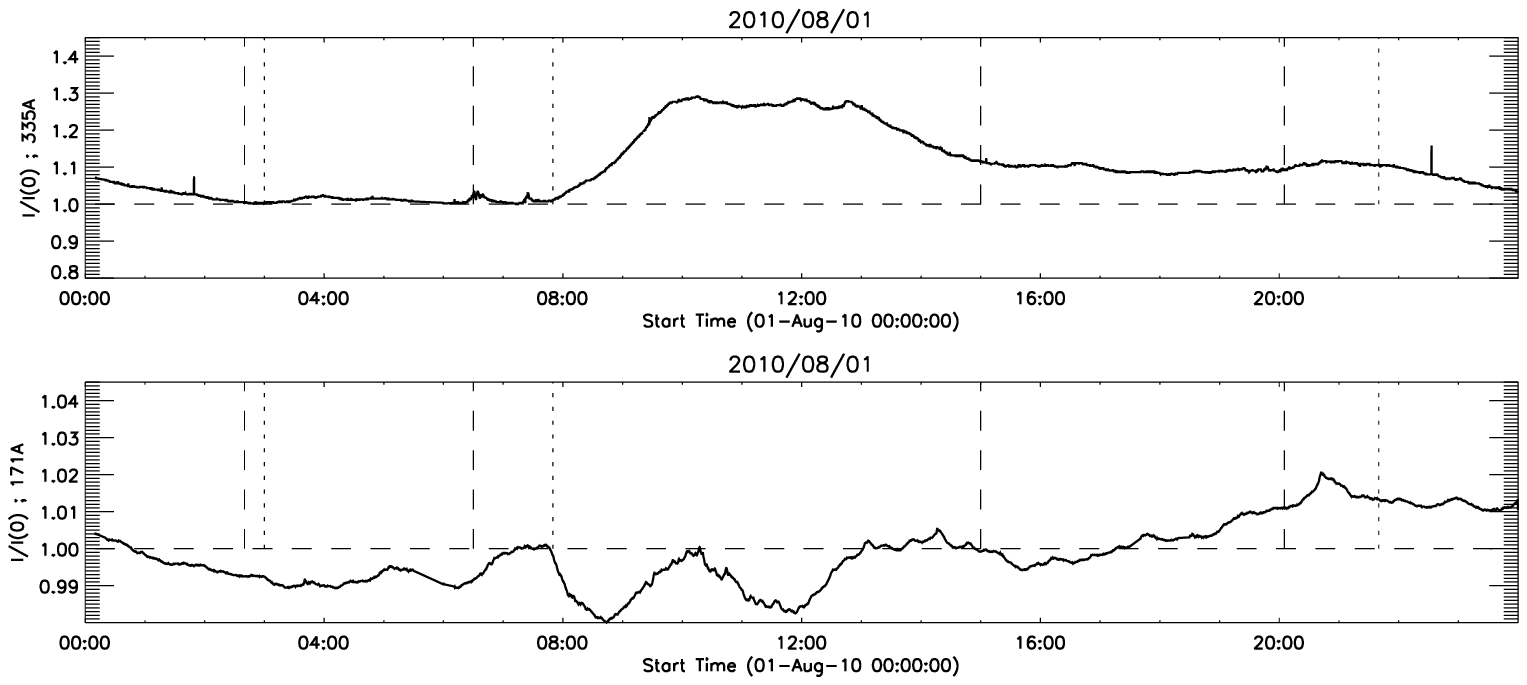}
\caption{Sample lightcurves observed with GOES (top) and with SDO/AIA in the Fe XVI
  335\,\AA\
channel (center) and the Fe IX/X 171\,\AA\ channel (bottom), for
a series of related coronal mass ejections and flares \citep[as
discussed by][]{schrijver+title2010}. The intensities were
scaled to the time of minimum brightness in the 335\,\AA\ channel 
(at the dashed line). Vertical dashed lines show approximate times of
the start of filament eruptions seen in AIA's 304\,\AA\ channel, and
dotted lines show when the first signs of CMEs were seen in STEREO's
COR1 images. \label{fig:lightcurves_20100801}}
\end{figure}
The right-hand panels in Fig.~\ref{fig:lightcurves} show the signals
in these same pass bands associated with the eruption of a large
quiet-Sun filament from an otherwise quiet-Sun disk. The X-ray
brightening barely registers as a flare, reaching no more than the B2
level. The Fe~XVI 335\,\AA\ signal shows a long-lived brightening
followed by a dimming before the signal recovers to pre-eruption
levels some 18\,h after the onset, while the Fe~IX/X 171\,\AA\ signal
persists as a brightening associated with quiet-Sun post-eruption
arcades for that entire period, starting about an hour after the onset
of the B2 flare. The total energy (fluence) associated with this event
(estimated from EVE measurements by R.\ Hock, priv.\ comm.) is $\sim
2\times 10^{30}$\,ergs, i.e. comparable to a low M-class flare, not dissimilar from that shown on the left. Despite the comparable energies involved in
these two events, the softness of the spectrum and the duration of this
very extended event, spanning a good fraction of a solar radius, lead
to an entirely different observational appearance if observed through lightcurves as can be obtained for stars.

\section{Local and global influences on eruptions}
High-resolution solar instruments typically have a limited
field of view, while telemetry constraints on, e.g., the full-Sun
X-ray imager on YOHKOH often led observers to down-select regions
around a flare site to bring down higher-cadence imaging. From
that perspective, it is not surprising that many studies looking for
conditions leading to the explosive/eruptive release of magnetic
energy focus on the conditions within or nearby an active region.
This has led to the recognition of certain patterns in the magnetic
field generally associated with flaring regions, in particular
neutral lines, often with high shear or high gradients
\citep[as summarized, e.g., by][]{schrijver2009b}. 

The capabilities of SDO's AIA, particularly when combined with EUV
observations from the two STEREO spacecraft that both approach
quadrature relative to the Sun-Earth line, are revealing that
long-range interactions play an important part in flares and eruptions
in addition to the 'internal conditioning'.
\citet{schrijver+title2010} discuss one particular set of flares and
eruptions in detail, but many more similar connections have been
inferred from other observations. They use full-sphere magnetic field
maps in combination with the STEREO and SDO observations to show that
a series of events on 2010/08/01 are directly connected by magnetic
field, in particular by field lines that are part of a web of
topological fault zones, i.e., separatrices, separators, and
quasi-separatrix layers. These long-range connections span across half
a hemisphere, part of which was invisible from Earth, but covered by
one of the STEREO spacecraft.

These observations reveal that in many cases, what may look like a
single event is instead a complex of simultaneous events, or events
that occur in close succession, across a large area of a stellar
surface. The solar events of 2010/08/01 are dominated by short
active-region flare loops in the X-ray part of the spectrum and by
long post-eruption loops over quiet-Sun filaments; the combination of
these wavelengths to estimate the properties of the field involved, as
one might do if these were stellar observations, would lead to
fundamentally erroneous conclusions. This is illustrated by the
lightcurves in Fig.~\ref{fig:lightcurves_20100801} (showing the same
set of signals as Fig.~\ref{fig:lightcurves}): the flares are
identifiable in the GOES X-ray signal (top panel), but already of a
substantially different character in the Fe~XVI 335\,\AA\ passband
shown below it. Around 1\,MK, in the Fe~IX/X 171\,\AA\ channel, the
flares are absent, while the dimmings are not obviously 
connected to the filament eruptions. Where
multi-wavelength coverage would help interpret stellar observations,
separation of the multitude of events on that day using only light
curves and then estimating total energies from available limited pass
bands is obviously an exercise fraught with large ambiguities and
uncertainties.  The repercussions on such linked (or ``sympathetic')
events on the hotly-debated problem of the causal links between flares
and CMEs continue to be studied; we may need to follow
\cite{harrison1996} who acknowledges the complex linkages between
these event classes and suggests that we refer to their composites as
'coronal magnetic storms,' a term equally suitable for the
stellar arena \citep[where sympathetic flaring is discussed, e.g., by][]{osten+brown1999}.

\begin{figure}[!ht]
%\plotone{flaredist.ps}
%\plotfiddle{flaredist.eps}{11cm}{0}{80}{80}{-230}{-30}
\plotfiddle{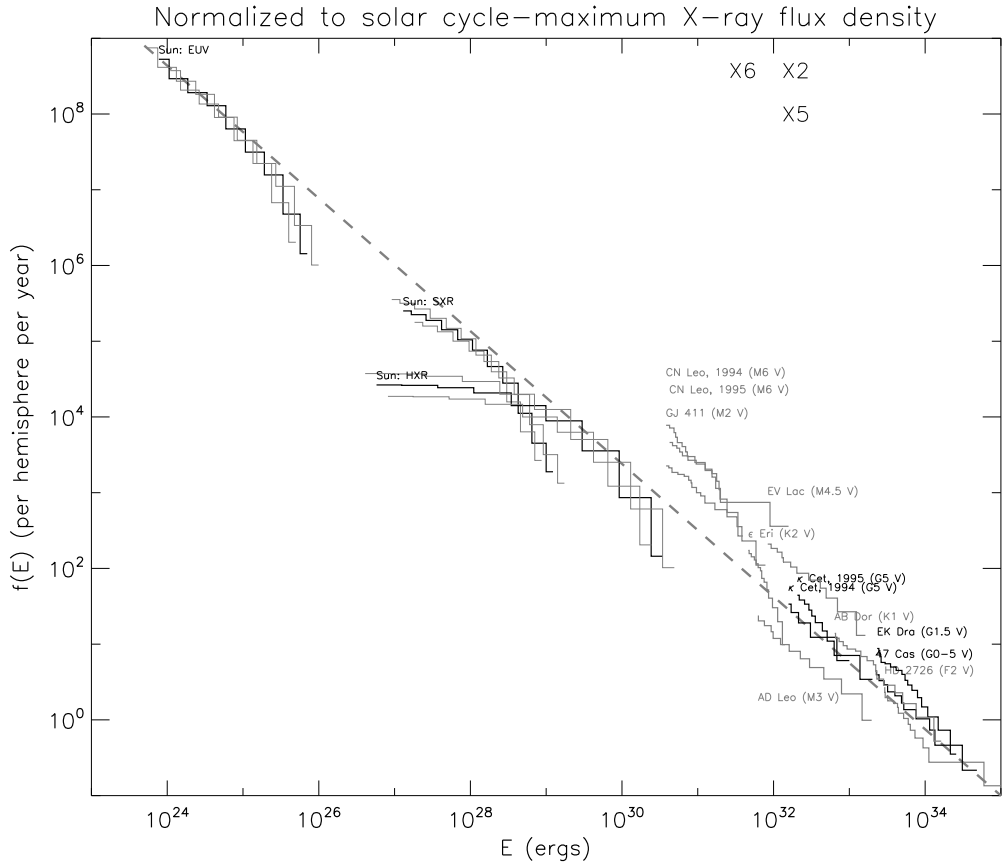}{11cm}{0}{80}{80}{-230}{-30}
\caption{Downward-cumulative flare frequency distribution for energies   
  exceeding $E$, 
  normalized to approximate solar-maximum X-ray flux density levels ($1.3\times
  10^5$\,erg cm$^{-2}$ s$^{-1}$; \citet{judge+etal2003f}) assuming a linear
  dependence of the X-ray surface flux density. Flare spectra for
  Sun-like G-type stars are shown in black, for warmer and cooler stars in
  grey. The grey dashed power-law fit has an index of $\alpha_f+1=-0.87$. EUV
  data from \citet{aschwanden99}, soft X-ray data from
  \citet{shimizu1997}, hard X-ray data $>8$\,keV
  from \citet{lin+etal2001a}, and stellar data from
  \citet{audard+etal2000a}.  The grey histograms for solar data
  bracket a conservative energy uncertainty of a factor of 2.  Three
  estimates of flare energies for GOES X flares are shown near the
  top, from \cite{aschwanden+alexander2001} and
  \cite{benz2008}. \label{fig:flaredist}}
\end{figure}
 \section{Flare energy spectrum}
Despite the difficulties that arise in establishing total energies
for the variety of flares even for solar events, the comparison from
nanoflares to X-class flares has been made, with the intriguing
result that flares span a range of at least 8 orders of magnitude
with a frequency distribution that can be approximated by a power-law
spectrum.

Statistics on stellar data are even harder to assemble and interpret,
but over the years
lessons have been learned and numbers extracted. One such study, by
\citet{audard+etal2000a}, assembles EUVE observations on 10 cool stars
into downward-cumulative frequency distributions (to suppress the
problem of low-number statistics) of estimated energies. Here, too,
power-law spectra arise, in fact with slopes very much like those
reported for solar flares. In another study, \citet{wolk_etal2005} look at young, active 
stars in the
Orion Nebula Cluster, just a few million years old; their flare
energy distribution yield
a power-law index of $\alpha_f\approx 1.7$.

\citet{osten+brown1999} --~for over 140\,d of observations on 16 tidally interacting RS\,CVn binaries~--
and \citet{audard+etal2000a} --~for some 74\,d of observing of 10 single and binary
stars~-- both conclude that the
overall flaring rate increases essentially linearly with the
background stellar X-ray luminosity, with flares reaching energies of
over $10^{34}$\,ergs, roughly comparable to a solar X500 flare. If
we use the inferred proportionality of flare frequency with X-ray
brightness to normalize the observed stellar frequency distribution
for flare energies to the Sun characteristic of cycle maximum, the
combined solar and stellar flare statistics (Fig.~\ref{fig:flaredist},
which shows the downward-cumulative frequency distribution) reveals a
frequency distribution for flare energies of
\begin{equation}\label{eq:freqdist}
N(E_f) {\rm d} E_f \propto E_f^{-\alpha_f} {\rm d} E_f,
\end{equation}
with $\alpha_f=-1.87\pm 0.10$ (rough uncertainty estimate). 
The solar data align remarkably well with those of the G-type
main-sequence stars in the sample by \citet{audard+etal2000a}.

One property of note here is that the slopes $\alpha_f$ in Eq.~(\ref{eq:freqdist}) 
and $\mu$ in Eq.~(\ref{eq:freqdistflux}) agree within their
uncertainties. What can we infer from this? 
The scaling of the flare frequency spectra observed for stars is
essentially linear with the X-ray brightness of the stars \citep{audard+etal2000a}, which,
in turn, is essentially linearly dependent on the unsigned magnetic flux
threading the stellar surface \citep{schrijver2000}, and modeling with
a surface flux-transport model suggests that that scales linearly with
the rate of emergence of bipolar active regions
\citep{schrijver2000}. 

With this chain of linear scalings, it is very tempting to conclude
that the flare frequency is proportional to the emergence frequency of
bipolar regions. One possible inference from this is that flares draw
their power from the same source as the active region magnetic
configurations.  In that context, it is interesting to note that the
flare energy distribution in Fig.~\ref{fig:flaredist} does not suggest
a cutoff in energy up to values approaching $10^{35}$\,ergs, which is
comfortably below the maximum available energy $E_{B,{\rm max}}$
estimated above for historically observed solar active regions.

Interestingly, recent Kepler observations of white light flares
reported at the 16th workshop on ``Cool Stars, Stellar Systems, and the
Sun''  suggest flares can occur up to at least $10^{37}$\,ergs, although
these most energetic flares are (at least at present) reported for
stars significantly cooler than the Sun, see, e.g.,
\citet{walkowicz+etal2010}.  What does it take to power such large
flares?  If we assume that a fraction of $f=0.01-0.1$ of the magnetic
energy density in a volume with a characteristic mean field strength
of $B_0$ can be converted into what eventually is radiated from the
flare site, the typical dimension $d_0$ and magnetic flux $\Phi_0=B_0
d_0^2$ in such a flaring region are given by
\begin{equation}
d_0=\left ( {4\pi E_f/f \over B_0^2} \right)^{1/3} \,\,;\,\, \Phi_o=B_0 d_0^2
= {(4\pi E_f/f)^{2/3}\over B_o^{1/3}}. 
\end{equation}
At $B_0=300$\,G, for flares with $E_f=10^{35}$\,ergs, $d_0\approx
(0.7-1.6)R_\odot$ and $\Phi_0\approx (0.8-4)\,10^{24}$\,Mx. Although
very sizeable, these numbers are still compatible with the largest
fluxes, $\Phi_{\rm max}$, discussed in Sect.~\ref{sec:bipoles}; note
that the value of $B_0$ here was chosen 3 times higher than
characteristic of solar regions to illustrate how challenging it is to
fit the flaring region on the available solar surface
\citep[see also, e.g.,][for evidence of extended flaring regions in some RS\,CVn stars]{osten+brown1999}. For
$E_f=10^{37}$\,ergs, on the other hand, $d_0\approx (3-7)R_\odot$ and
$\Phi_0\approx (20-80)\,10^{24}$\,Mx, which simply would not fit on
the Sun, and involves fluxes well above $\Phi_{\rm max}$; conditions
for such stellar flares must differ from those on the present-day Sun,
and perhaps --~as I discuss next~-- these very extreme events no longer
occur on the aged star next door.

\section{Spectrum of solar energetic particle events}
The frequency distribution in Fig.~\ref{fig:flaredist} suggests that
if we estimate an X1 flare to have an energy of about $10^{32}$\,ergs,
we should see about 30 flares per year of that magnitude or larger
during active phases in the solar cycle. That number compares quite
well with the average frequency of 25 per year for X-class flares over
the past three solar cycles, when counting only during the active half
of the cycles \citep[see the compilation by][]{hudson2007}. We would
expect an X10 or larger about 4 times per year, which is high by a
factor of about three given the 21 observed X-class flares since 1976
with the Sun in an active state for about 15 years within that
interval. Extending that spectrum even further, we would expect an
X100 flare or larger once every other year for the Sun near cycle
maximum.  As we have not experienced such large flares in at least
half a century, we face the possibility that despite the intriguing
alignment of solar and stellar flare data after normalization to the
mean coronal brightness, there may in fact be an upper limit to
stellar flare energies that may shift to lower values with increasing
age even as the spectrum below that value is left unchanged in slope:
the solar flare energy distribution appears to drop
below the power-law fit in Fig.~\ref{fig:flaredist} above X10, and the
largest flare observed to date has been estimated to be an X45$\pm 5$
\citep{thomson+etal2004}.

Are there constraints that can be set on the occurrence of
extreme flares that can be found in 'archives' on Earth other than
those compiled by mankind?  Solar eruptive events are frequently
associated with solar energetic particle (SEP) events that modulate
the galactic cosmic ray (GCR) background which itself varies on time
scales of years and longer
\citep{schrijver+siscoe2010b,schrijver+siscoe2010a}.  SEP events can
accelerate particles to energies up to several GeV which suffices to
initiate a nuclear cascade in the upper layers of the Earth's
atmosphere and in lunar or meteoric rocks. Some constraints on large
SEP events can be recovered by studying the mix of radioactive
nuclides or in
radioactive decay products as a function of penetration depth in lunar
and meteoric rock samples, or by studying the so-called odd nitrogen
concentrations that are modulated by reactions high in the Earth's
atmosphere that are sensitive to ionization states. The combination of
space-age measurements with such geological archival information led
\cite{usoskin_2008} to conclude that the downward-cumulative
probabilities of large solar proton events scaled with the fluence
$F_{\rm p}$ as a power law with a slope of $\delta+1 \sim 0.4$ for
relatively small events, then turning towards a steeper slope with a
lower limit of $\delta+1 \le 0.9$, with the 'break' occurring around
fluences of $10^{10}$\,cm$^{-2}$ for protons above 10\,MeV.

One might deduce from the inferred break in the SEP fluence
distribution that the behavior of solar flares changes at high
energies. That conclusion is ambiguous based on
this argument only, however, as argued by combining several
observed frequency distributions. Let us start from the particle
fluence distribution
\begin{equation}\label{eq:fluencedist}
N_{\rm p} {\rm d}F_{\rm p} \propto F_{\rm p}^{-\delta} {\rm d}  F_{\rm p} ,
\end{equation}
with the above value for $\delta$ as estimated from the cumulative
distribution function of the fluences. In order to relate the fluences
at Earth to those originating from the solar flares and associated
coronal mass ejections (both of which appear to contribute to the
SEPs in a mixture that continues to be debated), let us assume that
the particles are emitted from their source region in the corona
or inner heliosphere into a solid angle that scales with the total energy
$E_{\rm f}$ of the eruptive event as
\begin{equation}\label{eq:eruptangle}
\Omega \propto E_{\rm f}^\gamma .
\end{equation}
The value of $\gamma$ can be estimated by comparing the flare 
energy distribution in Eq.~(\ref{eq:freqdist}) and the distribution
of opening angles, $a$ (in degrees), for eruptions from very large CMEs seen 
by SOHO's LASCO to
small fibril eruptions observed by TRACE and perhaps for even smaller events seen in STEREO data \citep[summarized by][]{schrijver2010}
\begin{equation}\label{eq:angledist}
N_a {\rm d}a = b\, a^{-\beta} {\rm d a},
\end{equation}
with $\beta=2.0 \pm 0.3$ (estimated from Fig.~2
in \citet{schrijver2010}; $b\approx 1.1$ for $\beta =2$). 
For given $a$ (in radians), the corresponding fractional solid angle
is given by 
\begin{equation}\label{eq:openingangle}
{\Omega \over 4\pi} = {1 \over 2} ( 1- \cos{a}) \approx {1 \over 4} a^2,
\end{equation}
where the first term in the Taylor expansion shown on the right approximates
the solid angle only for sufficiently small values of $a$, otherwise
saturating when the hemisphere above the event is filled at $\Omega =
2\pi$. Using the right-hand expression in Eq.~(\ref{eq:openingangle})
together with Eq.~(\ref{eq:eruptangle}),
we 
can rewrite Eq.~(\ref{eq:angledist}) as 
\begin{equation}\label{eq:solidangledist}
N_a {\rm d}a \propto \Omega^{-{1 \over 2}(1+\beta)} {\rm d}\Omega
\propto E_f^{-{\gamma \over 2}(1+\beta)+\gamma -1} {\rm d}E_f.
\end{equation}
With Eq.~(\ref{eq:freqdist}), $\alpha_f ={\gamma \over 2}(1+\beta)-\gamma +1$ one
finds $\gamma = 2(\alpha_f-1)/(\beta-1)$.

Let us further assume that the particle fluence at Earth, $F_p$ is
a fixed fraction $f$ of $E_{\rm f}$, diluted
by
expanding over a solid angle $\Omega$, i.e., that with Eq.~(\ref{eq:eruptangle}),
\begin{equation}\label{eq:}
F_p \propto f {E_{\rm f} \over \Omega} \propto E_{\rm f}^{1-\gamma}.
\end{equation}
With this, Eq.~(\ref{eq:fluencedist}) can be transformed to read
\begin{equation}\label{eq:fluencedist2}
N_{\rm p} {\rm d}F_{\rm p} \propto E_{\rm f}^{-\delta
  (1-\gamma)-\gamma} {\rm d}E_{\rm f} \propto E_f^{\rm -\delta + 2
  (\alpha-1)(\delta -1)/(\beta-1)} {\rm d}E_{\rm f}.
\end{equation}
The assumption that the SEPs are emitted within a solid angle $\Omega$
implies that only those for which the direction of the Earth (mapped
through the curved path of the Parker spiral of the heliospheric
magnetic field) is included within that solid angle can be detected
(here, I ignore the fact that relatively small flares often do not
connect to the heliosphere, see Sect.~\ref{sec:spectral}). This means
that only a fraction
\begin{equation}\label{eq:probability}
p = {\Omega \over 4\pi} \propto E_{\rm f}^\gamma 
\end{equation}
of the total number of events is detected. Hence, to recover the flare
energy distribution from the observed SEP fluences, the distribution
in Eq.~(\ref{eq:fluencedist2}) has to be divided by $p$:
\begin{equation}\label{eq:fluencedist3}
N_f {\rm d}E_{\rm f} \propto E_{\rm f}^{-\delta + 2
  (\alpha-1)(\delta -2)/(\beta-1)} {\rm d}E_{\rm f} \equiv E_{\rm
  f}^{-\epsilon}  {\rm d}E_{\rm f}
\stackrel{?}{=}E_{\rm f}^{-\alpha_{f}}{\rm d}E_{\rm f}.
\end{equation}
With the values of the exponents above, one finds $\epsilon = 2.4 \pm
0.5$, consistent with the value of $\alpha_f$ in the flare
energy distribution of Eq.~(\ref{eq:freqdist}) within $\sim
1.1\sigma$, provided we limit the comparison to events for which the
SEPs are spread over a solid angle $a$ small compared to $2\pi$
steradians. This slope holds up to an SEP frequency at Earth of about 1/yr;
if we assume for those events $\Omega\approx 2\pi$, $p\approx 1/2$, 
so $\approx 2$/yr for the full Sun; Eq.~(\ref{eq:angledist}) has that downward 
cumulative frequency for $a\approx 200^\circ$, which is slightly high,
but roughly consistent with our hypothetical argumentation.
For relatively large opening angles $a$, the expression in
Eq.~(\ref{eq:openingangle}) saturates to a constant value, as does the
detection probability $p$. In that case, one would expect $\epsilon
\approx \delta \approx \alpha$. Such a value is just allowed by the
empirical upper limits to the energy distributions from Moon rocks. On
the other hand, such a slope may be too shallow relative to what is
suggested around the 'break' in the spectrum from NO$_3$ data from ice
cores as summarized by \cite{usoskin_2008}.

From this, we can tentatively conclude that at least the frequency
distributions for solar flares, for CME opening angles, and for SEP
fluences are consistent for proton events up to the largest observed
in the instrumental era. The break in the fluence spectrum above those
values might reflect the saturation of the spreading of the SEPs over
essentially a full hemisphere over the solar source region. But it is
possible that a true saturation occurs somewhere along the chain of
events from Sun to Earth: flare energies may have an upper cutoff for
the present-day Sun (with flare probabilities dropping significantly
below the power-law fit in Fig.~\ref{fig:flaredist} for flares above
X10) that is not readily inferred by looking at samples of young Suns
or the generation of energetic particles may be limited (either at the
flare site or within heliospheric shocks).

\section{In conclusion}
The sample observations discussed above demonstrate that the
combination of geophysical, heliophysical, and astrophysical data can
teach us much about the Sun's magnetic climate, up to the most
energetic of events. The interpretations outlined are, of course, to
be tested and alternatives, that doubtlessly exist, are to be
explored. Despite the speculative nature of the scenarios sketched above, 
it appears that we are close to having
the material available to learn where the solar flare-energy spectrum
drops below a solar-stellar power law: the combination of the study of
archives in ice and of stellar flare statistics should be able to
provide us with an answer. On a less positive note, the solar
observations discussed above demonstrate that measuring the energies
involved in explosive and eruptive events is difficult, that
separating events based on lightcurves alone is an
ambiguous exercise, and that broad wavelength coverage is essential to
both of these objectives: to learn about the most severe space
weather, we have to accept that long-duration, large-sample,
pan-chromatic (and thus often multi-observatory) stellar observations
are needed because they can provide crucial information that can
otherwise only be gathered by observing the Sun for a very long time and undergoing the detrimental effects of extreme magnetic storms on the Sun and around the Earth.

\acknowledgements I thank M.\ Aschwanden, J.\ Beer, R.\ Osten, 
C.\ Parnell, and A.\ Title for discussions, suggestions, and 
pointers to the literature.

%\bibliography{/Users/schryver/book/references/ref_karel}

\end{document}